\documentstyle[prl,twocolumn,aps,epsf,epsfig]{revtex}
\begin{document}
\draft
\twocolumn[\hsize\textwidth\columnwidth\hsize\csname@twocolumnfalse%
\endcsname

\title{Composite Spin Waves, Quasi-Particles and
Low Temperature resistivity in  Double Exchange Systems. }

\author{M.J. Calder\'on and L. Brey}
\address{Instituto de Ciencia de Materiales de Madrid, 
CSIC,
28049 Cantoblanco, Madrid, Spain.}

\date{\today}

\maketitle

\begin{abstract}
\baselineskip=2.5ex
We make a quantum description of the  electron low temperature  properties of 
double exchange materials. In these
systems there is a strong coupling between the core spin and the carriers spin. This 
large coupling makes the low energy spin waves to be a combination of ion and electron
density spin waves. We study the form and dispersion of these 
composite spin wave excitations. We also analyze the spin 
up and down spectral functions of the temperature dependent quasi-particles
of this system. 
Finally we obtain  that the thermally activated composite spin waves renormalize 
the carriers
effective mass and this gives rise to a low temperature resistivity scaling as
$T ^{5/2}$.

\end{abstract}
]

Doped perovskite manganites have attracted much attention lately, since 
they 
undergo a ferromagnetic-paramagnetic transition accompanied
by a metal-insulator transition\cite{review}.
The Double-Exchange (DE) mechanism\cite{detheory} plays
a major role to explain this magnetic transition. In the DE picture,
the carriers moving through the lattice are strongly ferromagnetically coupled
to the Mn ion  spins producing a modulation of the hopping amplitude
between neighboring Mn ions.

A big effort has been done in understanding the electron  transport 
properties of these materials
at temperatures $T$ near the critical temperature\cite{mogo}.
However, the low $T$ properties of these systems are poorly understood. At low $T$
a dominant $T ^2$ increase in resistivity, $\rho$, is generally observed\cite{mogo2},
$\Delta \rho \sim T ^2$.
Although the $T^2$ behavior is similar to that produced by electron-electron interaction,
the coefficient of the $T ^2$ term is about 60 times larger than the expected for 
electron-electron scattering and therefore this mechanism has been ruled out.
Another source for this $T^2$ behavior is single spin wave scattering. However in DE 
materials only one spin channel is metallic and single  spin wave
scattering process is prohibited. Two spin wave scattering gives a $T^{9/2}$ 
increase\cite{kubo},
clearly slower than the experimental data.
In ref\cite{g-m} the temperature dependence of the resistivity is attributed
to polaron coherent motion. At low T this process gives  $\Delta \rho \sim T ^ 2$ 
and a good fit of the resistivity is obtained.

Furukawa \cite{furukawa} propose an unconventional one-magnon scattering in half metals,
which gives $\Delta \rho \sim T ^3$. We believe this dependence is not right, 
because they calculated the inverse of the imaginary part of the electron self-energy
and not the transport scattering time; an appropriated  calculation
taking into account the fractional loss of forward velocity\cite{1mcos}
will give a $T^{7/2}$ dependence of the resistivity. On the other hand, the one magnon scattering
is calculated in first order perturbation theory in the Hund's coupling, $J_H$, between the
electron and ion spins. This coupling is assumed to be infinity in the DE model
so that perturbation theory in this parameter is not valid (in particular it would imply
a zero lifetime for the carriers).
Wang and Zhang \cite{wang} assume that the minority spin electron states are
localized and obtain  $\Delta \rho \sim T ^ {3/2}$, however again in this approach 
the scattering time is proportional to $J_H$.

In this work we analyze some low $T$ properties of DE materials. We study the low energy
spin waves and we find that, in order to minimize the energy, the ion spin waves (ISW) and
the electron  spin density waves (ESDW) become coupled forming a composite
spin wave (CSW) with energy independent of $J_H$. 
By analyzing the Hamiltonian of the system in presence of CSWs we obtain the spin up and down
spectral weights of the $T$ dependent quasi-particles.
These  spectral functions do not depend on $J_H$. Finally we analyze the low $T$ resistivity
of the DE systems. We obtain that the main effect of $T$ is to  renormalize
the carriers effective mass, $m ^* $, which implies an increase of
the resistivity  $\Delta \rho \sim T ^ {5/2}$

{\it Composite Spin Waves.}
The Hamiltonian describing the Mn oxides is, 
\begin{equation}
\hat H    =    -  t \sum _{i \ne j, \sigma} 
C ^+ _{i,\sigma}  C  _{j,\sigma} - 
{{ J_H} \over {S}}  \sum _{i,\sigma,\sigma '} C ^+ _{i,\sigma} 
{\mbox {\boldmath $\sigma$}} _{\sigma,\sigma'}
C_{i,\sigma'} {\bf S}_i  \, \, \, ,
\label{hamil1}
\end{equation}
here $ C ^+ _{i,\sigma}$ creates an electron at site $i$ and with spin $\sigma$,
$t$ is the hopping amplitude between nearest-neighbor sites,  
and 
${\bf S}_i$ is the ion  spin at site $i$. 
The ions are assumed to be located on single cubic lattice with  lattice parameter $a_0$.
The second term in Hamiltonian (\ref{hamil1}) describes the ferromagnetic 
coupling between the electron and ion spins. Because of this coupling the electron spins 
prefer to be 
parallel to the ion spins. The electrons can lower their  energy by hopping
from site to site.  
And 
in order to minimize 
the electron kinetic energy, the ion spins become ferromagnetically coupled.
The ground state (GS) of the system is half metallic in the sense that the system is 
metallic for one spin orientation but it is insulator for the opposite orientation.

Writing the  electrons operators as Bloch operators, $C^+ _{ {\bf k} \sigma} $, 
and representing the ion  spins in term of Holstein-Primakoff bosons\cite{mattis}, 
in first order in the $1/S$ expansion 
the Hamiltonian get the form:
\begin{eqnarray}
H & = &  \sum _{ {\bf k} \sigma } \varepsilon _{\bf k} 
C^+ _{ {\bf k} \sigma} C _{ {\bf k} \sigma}  - J_H \sum _ { {\bf k} \sigma }
\sigma \,  C^+ _{ {\bf k} \sigma} C _{ {\bf k} \sigma} \nonumber \\
& -& J_H \sqrt{ {2 \over {S  N}}} 
\sum _{ {\bf q}, {\bf k}} \left ( 
b _q ^+ \, C^+ _ { {\bf k} \uparrow} C _{ {\bf k} + 
{\bf q}  \downarrow} 
+
b _q  \, C^+ _ { {\bf k} + {\bf q}  \downarrow} 
C _ { {\bf k} \uparrow} \right ) 
\nonumber \\
& + & { J_H \over {N  S}} \sum _{ {\bf k}, {\bf q} _ 1 , {\bf q}_2, \sigma }
\sigma \, b ^ +  _ { {\bf q }_1}
b  _ { {\bf q }_2}
C ^+ _ { {\bf k}- {\bf q}_1 \sigma}
C  _ { {\bf k}- {\bf q}_2 \sigma}
\, \, \, ,
\label{hamil2}
\end{eqnarray}
here $N$ is the number of sites in the system, $\varepsilon _ {{\bf k}}
= - 2 t \sum _{\alpha} \cos{(  {k_{\alpha} } a_0)} $, is the electron energy spectrum
and 
$b^+ _{\bf q}$  creates a ISW 
with momentum ${\bf q}$, which 
decreases the $z$-component of the total ion  spin by unity.
In the above Hamiltonian the sum of momenta is restricted to the 
first Brillouin zone.
In the DE case ($J_H \rightarrow \infty$), 
the GS of this Hamiltonian is ferromagnetic, 
with all the electron spins and core spins parallel, let us say pointing up in the 
$z$-direction.
The ground state energy per electron is $E_0=-J_H + E_{KE}$, being
$E_{KE}= {1 \over {N_e}}\sum _{\bf k} ^{occ} \varepsilon _{\bf k}$
the average kinetic energy per electron, 
$N_e$ the number of electrons and  the sum is over the occupied electronic states.

We are interested in the low energy ($\sim t$)
spin excitations of the Hamiltonian (\ref{hamil2}). A rotation 
of an  electron spin 
costs an energy of the order of $J_H$, and there are not
low-lying single particle spin  excitations. Also the creation of an 
ISW costs an energy of the order of $J_H$. Therefore the only 
low-energy spin excitations are collective modes, 
in which the energy is minimized by coherently distributing the 
momentum and the spin loss among a large number of electrons and core
spins. 
Furthermore, as in the case of the single
mode approximation developed by Feynman for helium\cite{feynman}, 
the absence of low-energy electron spin flip excitations
makes that these collective excitations  remain well defined
even for microscopic  wavelengths.

In the spirit of the single mode approximation, we expect the low energy
mode to be a linear combination of an ISW  and 
an ESDW defined by the bosonic
operator
\begin{equation}
a^+ _{\bf q} = { 1 \over {\sqrt {N_e}}} \sum _{\bf k} 
C^+ _{ {\bf k} + {\bf q} \downarrow} C_{{\bf k} \uparrow} \, \, \, .
\end{equation}
The form and the  energy of the excitation are obtained 
by diagonalizing the matrix
\begin{equation}
\left ( 
\begin{array} {cc}
\, \, \, \, \,  <|a_{\bf q} [ H, a_{\bf q} ^+ ] |> \, \, \, \, \, & 
\, \, \, \, \, <|b_{\bf q} [ H, a_{\bf q} ^+ ] |> \, \, \, \, \, \\
\, \, \, \, \, <|a_{\bf q} [ H, b_{\bf q} ^+ ] |> \, \, \, \, \, & 
\, \, \, \, \, <|b_{\bf q} [ H, b_{\bf q} ^+ ] |> \, \, \, \, \, 
\end{array}
\right )
\end{equation}
where the expectation value is obtained in the ferromagnetic GS.
Using  Hamiltonian (\ref{hamil2}) this matrix becomes,
\begin{equation}
\left ( 
\begin{array} {cc}
\, \, \, \, \, 2 { {NS } \over {N_e}} \omega ({\bf q}) + 2 J_H \, \, \, \, \,  &
\, \, \, \, \,  - J _H  \sqrt { {{ 2 N_e} \over {NS}}}\, \, \, \, \,  \\
\, \, \, \, \, - J _H  \sqrt { {{ 2 N_e} \over {NS}}}\, \, \, \, \,  & 
\, \, \, \, \, J_H { {N_e} \over {NS}} \, \, \, \, \, 
\end{array}
\right )
\end{equation}
Here 
\begin{equation}
\omega ({\bf q})= -{ {E_{KE}} \over {3S}} 
\sum _{\alpha} \sin ^2 \left ( { {q_{\alpha} a_0} \over 2} \right ) 
\, \, \, .
\label{wq}
\end{equation}
In the $J_H \rightarrow \infty$ limit the eigenvectors for this matrix are 
\begin{eqnarray}
\Lambda ^+ ( {\bf q})  & = &  b_{\bf q} ^+ + \sqrt{{ {N_e} \over {2SN}}} a_{\bf q} ^+ \\
\Xi ^+ ( {\bf q})  & = & \sqrt{{ {N_e} \over {2SN}}}    b_{\bf q} ^+  -  a_{\bf q} ^+ 
\end{eqnarray}
which correspond  to the following energies,
\begin{eqnarray}
\omega _1  ( {\bf q})  & = & \omega ({\bf q})  \\
\omega _2  ( {\bf q})  & = & 2 J_H + J_H { { N_e } \over {NS}} + 2 {{NS} \over {N_e}} \omega
({\bf q})  
\end{eqnarray}
For comparison with the phonons these two modes represent acoustic and optical  spin waves.
$\omega _1  ( {\bf q})$, see Eq.\ref{wq}, is proportional to $t$, and does not depend on the
Hund's coupling $J_H$. 
At long-wavelengths $\omega _1  ( {\bf q}) $=$\rho_s q ^2$,
being $\rho_s = - E_{KE}/(12S) a_0 ^2$ the spin stiffnes.
$\omega _1  ( {\bf q})$ is a gapless Goldstone mode 
reflecting the spontaneous breaking of rotational symmetry.
Its quadratic dispersion in ${\bf q}$ reflexes the O(3) symmetry
of the underline Hamiltonian.

The expression obtained for the low energy mode is equal to that  obtained previously by using
second order  perturbation theory in $1/S$\cite{furu1,nagaev,golosov}.
The operator $\Lambda ^+ ( {\bf q})$ acting on the ferromagnetic GS creates 
a symmetric combination
of an ISW and an ESDW, in such a way that at each place the expectation 
value of the core spin and the electron spin are parallel; this is the reason 
why the energy scale of this excitation is $t$.
We call the excitation created by $\Lambda ^+ ( {\bf q})$ a composite spin wave (CSW).

The operator $\Xi ^+ ( {\bf q}) $ creates an antisymmetric combination of 
an ISW and an ESDW, this collective mode has an energy above the Stoner continuum.

These two modes,  $\Lambda ^+ ( {\bf q})$  and $\Xi ^+ ( {\bf q}) $ 
are equivalent to those found in diluted magnetic semiconductors\cite{konig}. 
Note, however, that in semiconductors the coupling between the carriers and the
Mn is antiferromagnetic and the high energy mode occurs at an energy below the
Stoner continuum.

{\it Finite temperature quasi-particles.}
Now we study how the electronic properties  are modified
by the presence of thermally activated CSWs.
In order to do that we consider 
the system in presence of a static and semiclassical CSW and we get  the 
electron energy spectrum,  wave function and  Green function 
in presence of this perturbation.
After quantizing  the spin waves, we obtain the finite temperature 
quasi-particles in the system.

Semiclassically the presence of a CSW of wave-vector ${\bf q}$, reduces  the 
$z$-component of all the ion spins from $S$ to $S\cos \theta$, and 
the transverse components of an ion spin at site $i$ are given by
$S_x+ i S_y = S \sin \theta \, e ^{i {\bf q} {\bf R} _i}$. 
At each site $i$ the two electron spin states
are the parallel (energy -$J_H$) and the antiparallel (energy $J_H$)
to the ion spin ${\bf S} _i$, 
and the low energy state at site $i$ is,
\begin{equation}
d^+ _ i = \sin { \left( { {\theta } / 2} \right ) } \,  
e ^{ - i {\bf q} {\bf R } _i} C ^+ _{i , \downarrow} -
\cos  { \left ( {\theta } / 2\right ) }  C ^+ _{i , \uparrow} 
\, \, \, .
\end{equation}
Note that in the above definition there is an implicit dependence of the 
$d^+$ operators on ${\bf q}$.
In the following we can forget about the high energy electron states which are
$2 J_H$ higher in energy.
The tunneling between two neighbors low-energy orbitals located 
at $i$ and $i+{\hat {\alpha}}$, is 
\begin{equation}
t ' _{i,i+{\hat {\alpha}}}
\equiv  t'_{\alpha} ({\bf q}) 
= t \left  ( \cos ^2 {\left ( { \theta } / 2 \right ) } 
+\sin  ^2 {\left ( { \theta } / 2 \right ) } \, e ^ {i q _{\alpha} a _0} \right) 
\end{equation}
which does not depend on site. 
The electron energy spectrum in presence of a CSW is,
\begin{equation}
\varepsilon ' _{\bf k} ( {\bf q})  = - 2 \sum _{\alpha} |t ' _{\alpha} ({\bf q})|\,  
\cos {k _{\alpha}a _0 } - J _ H \, \, . 
\end{equation}
and the 
low energy part of the Hamiltonian takes the form, 
$
{\hat H}  = \sum _{\bf k} \varepsilon ' _{\bf k} ({\bf q}) \, d ^+ _{\bf k} d _{\bf k} 
$.
Note that the eigenvalues of the system in presence of a CSW are defined
in the same Brillouin zone than those of  the ferromagnetic GS. 
However, the low energy electron basis states depend on position, and the presence of the  CSW
of momentum ${\bf q}$
implies the mixing of states of 
momentum ${\bf k}$ and ${\bf k} -{\bf q}$ and opposite spins; 
\begin{equation}
d ^+ _ {\bf k} = \sin \left ( {{\theta } / 2} \right ) 
C ^+ _{ {\bf k} - {\bf q} \downarrow} - 
\cos \left ( { {\theta } / 2} \right ) 
C ^+ _{ {\bf k}  \uparrow}  \, \, \, .
\end{equation}
The low energy part of the Green function takes the form,
\begin{eqnarray}
G ( {\bf k}, \omega )   =  & &   
\left ( \omega - \varepsilon _{\bf k} ' ({\bf q}) \right ) ^{-1} \times 
\nonumber \\
& & \left \{    \sin   ^2 \left ( {{\theta } / 2} \right )
C ^+ _{ {\bf k} - {\bf q} \downarrow} C  _{ {\bf k} - {\bf q} \downarrow} 
  +  \cos ^2 \left ( {{\theta } / 2} \right )
C ^+ _{ {\bf k}  \uparrow} C  _{ {\bf k}  \uparrow} \right . \nonumber \\
&   &  - \left . {{ \sin \theta} \over 2} ( C ^+ _{ {\bf k} - {\bf q} \downarrow} C  _{ {\bf k}  \uparrow} + C ^+ _{ {\bf k}  \uparrow} 
C  _{ {\bf k} - {\bf q} \downarrow} ) \right \}
\label{green}
\end{eqnarray}

Now we quantize the spin waves. 
The total spin in the zero temperature GS of the system is $S_T=NS+N_e/2$.
The values of the  total  spin  should be 
$S_T$, $S_T-1$, $S_T-2$..., and therefore in presence of  
spin waves the  
$z$-component of the total spin  should change in a integer number 
which represents the number of thermally activated CSW, $n_q $=$\Lambda ^+ _{\bf q}
\Lambda  _{\bf q}$. With this,  the quantization rule is 
$\cos \theta = 1 - {{ n_q} \over {S_T}} $ and  
the energy of the system associated with the  presence of 
thermally activated CSW is
\begin{equation}
E=\sum _{\bf q} \sum _{\bf k} ^{occ} \left ( \varepsilon _{\bf k} ' ( {\bf q}) -
\varepsilon _{\bf k} \right ) 
= \sum _{\bf q} n _{\bf q} \hbar \omega ({\bf q}) \, \, \, .
\end{equation}
From the above Green function and using the quantization rule, we can define
$T$ dependent quasi-particles in the system, with spectral weights in spin 
up and spin down electron states, 
\begin{eqnarray}
A _ {\uparrow} (\bf {k}, \omega ) & = &  {1 \over {\pi}} \, ( 1 - { { \delta m (T)  } \over 2}) \, \delta ( \omega -
\tilde {\varepsilon} _{\bf k} ) \nonumber \\ 
A _ {\downarrow} (\bf {k}, \omega ) & = &  {1 \over {\pi}} \, { { \delta m ( T )  } \over 2} \, \delta ( \omega -
\tilde {\varepsilon} _{\bf k} ) \, \, \, \, 
\label{spectra}
\end{eqnarray}
In the above expressions
$
\delta m  ( T ) = 1 - {{ M ( T ) } \over {M (0)}} \sim T  ^{ 3 \over 2}
$ and 
\begin{equation} 
{{ M ( T ) } \over {M (0)}} = 1 - { 1 \over {M (0)}}  \sum _{\bf q} n _{\bf q} \,  \, 
\end{equation}
is the relative  magnetization suppression  due to thermal CSW excitations
and $\tilde{\varepsilon} _{\bf k} $ is the quasi-particle energy, 
\begin{equation}
\tilde{\varepsilon} _{\bf k} = \varepsilon _{\bf k} -
2 t \sum _{\bf q}   \left ( 
 {{ n _q }  \over { S_T }} \right ) \sum _{\alpha} 
\sin ^2 { { q _{\alpha} a _0 } \over 2} 
\cos k _\alpha a _0 
\end{equation}
As we will comment below the band-width of the system decreases with $T$.
In obtaining Eq. (\ref{spectra}) we have summed over all thermally activated CSW and we 
have assumed that at temperatures of interest the wave-vectors of the CSW are small
compared to the Brillouin zone dimensions.

From the spectral function we see that the quasi-particle, which has its spin aligned with
the fluctuating ion spins, will,  at finite $T$,  be a spin up  state
with probability 
$ (1 -\delta m (T )  / 2) $ and a spin down state with probability $\delta  m ( T )  / 2$.
The appearance of a spin down shadow band\cite{ahm} at energies $-J_H$ is due to the 
thermal excitation of low energy long-wavelength CSW.
The relative electron spin polarization and ion spin polarization have the same value
and 
scale as
$T^ {3/2}$.
Note that the spectral weight does not depend on  Hund's coupling $J_H$.

{\it Low temperature resistivity.}
A CSW 
modifies the value of the hopping amplitude
however its presence   does not
modulate spatially the value of $t$. 
As we commented above
the size of the Brillouin zone is not modified by the presence of a
CSW.
This implies that an  electron is not scattered by a single CSW; 
the  electron
creation operator evolves continuously from $ C ^ + _ {\bf k}$ to $ d ^+ _ {\bf k} $.
The reason for this behavior is that in 
the $J_H \rightarrow \infty $ limit, and in the adiabatic approximation, the 
electron spins follow instantaneously the core spin fluctuations.
The adiabatic approach is based in the fact that the core-spin fluctuates at frequencies
related only to the temperature which 
is assumed to be much smaller than the hopping amplitude.

Let us see this  more carefully. In the $J_H \rightarrow \infty $ limit,
the low energy part of the Hamiltonian (\ref{hamil1}) gets the form,  
\begin{equation}
H = - t \sum _{ i \ne j} { 1 \over S} \sqrt { {{ S ^2 + {\bf S } _ i {\bf S} _ j } \over 2 } }
\, d ^+ _ i \, d _ j 
-J_H \sum _ i d^ + _i d _i 
\, \, \, \, 
\label{hamilde}
\end{equation}
where $d_i ^+ $ creates
an electron at site $i$ with
its spin parallel to ${\bf S }_ i$.
Assuming small fluctuations of the core spins with respect to  the ferromagnetic GS,
the Hamiltonian (\ref{hamilde})  can be written in the form,
\begin{eqnarray}
H   =   \sum _{\bf k} ( \varepsilon _{\bf k}  -  J_H ) d _{\bf k } ^ +  d _ {\bf k}  
& + &  
{1\over {4S _ T }} \sum _ { {\bf k}, {\bf q}, {\bf q}' }
d _{\bf k} ^+ d _ {{\bf k}+{\bf  q}} \, b ^+ _ {{\bf q} + {\bf q}'}
b _{ {\bf q }' }
\nonumber \\
& \times & 
\left ( \varepsilon _{ {\bf k}-{\bf q}} 
+
\varepsilon _{ {\bf k}' +{\bf q}} 
-
\varepsilon _{ {\bf k}} 
-
\varepsilon _{ {\bf k}' } \right )
\end{eqnarray}
It is clear in this 
expression that there is not scattering
of an electron by a single CSW. The only source of scattering is a two CSW process,
which produces, in second order in $1/S$,  
an inverse scattering time proportional to $T ^{9/2}$ and therefore a 
low temperature resistance also proportional to 
$T ^ {9/2}$\cite{kubo}. 

The presence of CSWs also changes the electron self-energy. In this case, 
an electron with momentum ${\bf k}$ 
and spin locally parallel to the ion spins has in first order in $1/S$, 
a self-energy,
\begin{equation}
\Sigma _{\bf k} = {1 \over {2S_T}}
\sum _{\bf q}  n _ q \left ( \varepsilon _{ {\bf k} + {\bf q}} - \varepsilon _{\bf k} \right )
 \, \, \, .
\end{equation}
Note that $\Sigma _{\bf k}$ is equal to 
$
\tilde{\varepsilon} _{\bf k}$-$\varepsilon _{\bf k} 
$,
which is the self-energy obtained by quantizing the CSWs.
In the case of low $T$ the self-energy gets the form,
\begin{equation}
\Sigma _{\bf k} =
 -  { 1 \over {12}} { N \over {S_T}} C \left ( { {KT} \over {\rho _s}} \right ) ^ {5 \over 2}
\varepsilon _{\bf k} 
\, \, \,
\end{equation}
where $C= {1 \over {4 \pi ^2}} \int _0 ^ {\infty} 
{ { u ^{3/2}} \over {e ^ u -1 }} d u \approx 0.045 $ is a constant.
The thermally activated CSWs reduce the average transfer integral 
and lead to a decrease of the electrons band-width and therefore to a  renormalization  of the 
electron effective mass $m^* $. The effective mass increases as $m ^* \sim T ^{5/2}$.
The change in the band-width has a strong effect on the electronic transport
properties of the system; in real systems the presence of impurities
produces a resistivity $\rho$ given by the Drude formula,
\begin{equation}
\rho = { {m ^* } \over { n e ^2 \tau}} \, \, \, 
\end{equation}
where, $n$ is the electron density and $\tau$ is the relaxation time due to the presence
of imperfections in the system. The increase of  $m ^* $ with T temperature
implies an increase of $\rho$\cite{mass}. Therefore we obtain that in real systems
the resistance at low temperatures behaves as $T^{5/2}$.
The coefficient of this $T^{5/2}$ term is proportional to the zero temperature 
resistivity of the system, and its importance with respect to other terms will
depend on the quality of the sample.
The existence of spin-orbit coupling will induce a gap in the CSW spectrum so that  the effective
mass and the resistivity will remain practically constant up to a temperature
of the order of this gap, and then they will scale as $T^{5/2}$.

In conclusion we have studied the low energy and low $T$ electronic properties
of DE systems. We have obtained that the low energy spin excitations are composite spin waves:
a lineal combination of ion and electrons spin waves. We have also analyzed the spectral function
of the $T$ dependent quasi-particles. Finally we have obtained that the thermally activated 
spin waves renormalize the effective mass of the carriers and this produces
that the low $T$ resistivity of the system scales as $T ^{5/2}$

The authors thanks C.Tejedor, A.H.MacDonald and F.Guinea for
helpful discussions. This work was supported by the
CICyT of Spain under Contract No.PB96-0085 and by the CAM under contract
No. 07N/0027/1999.


\begin{references}


\bibitem{review}J.M.D. Coey {\it et al.}, Adv. Phys. {\bf 48}, 167
(1999); A.P. Ramirez, J. Phys. Condens. Matter {\bf 9}, 8171 (1997).
\bibitem{detheory}C. Zener, Phys. Rev. {\bf 82} 403 (1951);
P.W. Anderson and H. Hasegawa,{\it ibid.} {\bf 100}, 675 (1955);
P.G. deGennes, {\it ibid.} {\bf 118}, 141 (1960).
\bibitem{mogo}E.E.Narimanov {\it et al.}, cond-mat/0002191;
A.J.Millis {\it et al.} Phys.Rev.B, {\bf 54}, 5405 (1996);
V.N.Smolyaninova {\it et al.} Phys.Rev.B {\bf 62}, 3010 (2000);
M.J.Calder\'on {\it et al.} Phys.Rev.B {\bf 58}, 3286 (1998);
{\it ibid} {\bf 59}, 4170 (1999); Daniel P. Arovas {\it et al.} Phys.Rev.B {\bf 58}, 9150
(1998); S.K. Sarker, J. Phys.:Condens. Matter {\bf 8} L515 (1996).
\bibitem{mogo2}
M.Ziese, Phys.Rev.B {\bf 62}, 1044 (2000);
P.Schiffer {\it et al}, Phys. Rev. Lett.{\bf 75}, 3336 (1995)
M.Jaime {\it et al.} Phys.Rev.B {\bf 58}, R5901 (1998).  
\bibitem{kubo}K.Kubo {\it et al.} J.Phys.Soc.Jpn. {\bf 33}, 21 (1972).
\bibitem{g-m}G.-m Zhao {\it et al.} Phys.Rev.Lett. {\bf 84} 6086 (2000).
\bibitem{furukawa}N.Furukawa, J.Phys.Soc.Jpn. {\bf 69} 1954 (2000).
\bibitem{1mcos}That corresponds to considering  the 
$1-\cos {\theta}$ term. See for example W.A.Harrison, {\it Solid State Theory}
(Dover, NY 1980).
\bibitem{wang}X.Wang {\it et al.} Phys.Rev.Lett. {\bf 82}, 4276 (1999).
\bibitem{mattis}D.C.Mattis, {\it The Theory of Magnetism} (Springer, Berlin 1981).
\bibitem{feynman}R.P.Feynman, {\it Statistical Mechanics}
(Benjamin Reading Mass. 1972).
\bibitem{furu1}N.Furukawa, J.Phys.Soc.Jpn. {\bf 65}, 1174 (1996).
\bibitem{nagaev}E.L.Nagaev, Phys.Rev.B {\bf 58}, 827 (1998).
\bibitem{golosov}D.I.Golosov, Phys.Rev.Lett. {\bf 84}, 3974 (2000).
\bibitem{konig}J.Konig {\it et al.} Phys.Rev.Lett. {\bf 84}, 5628 (2000).
\bibitem{ahm}A.H.MacDonald, T.Jungwirth and M.Kasner, Phys.Rev.Lett. 
{\bf 81}, 705 (1998).
\bibitem{mass}Note that, as we are dealing with first order perturbation theory,
the self-energy does not depend on frequency but only on $\bf k$. This implies that
the mass is renormalized but the relaxation time is not. On the contrary, 
in electron-phonon interaction, both quantities are equally renormalized giving a 
resistivity independent of the mass enhancement (see R.E. Prange and L.P. Kadanoff,
Phys. Rev. {\bf 134}, A566 (1964)).
\end{references}
\end{document}